**Talks that Builds: Exploring Communication factors for the Success of Emerging Professional in Product Teams**

**Nyan Lin Zaw (Imran)**

**Berea College**

November 23$^{th}$, 2025

Abstract

This paper recognizes that most organizational communication study focuses on established professionals aged above 27 with more than five years of experience. In contrast, this study examines product teams with younger "emerging" professionals aged 18-27 and explores which factors influence their success. While some established factors still apply, others become less relevant, and new ones such as curiosity, locational proximity, documentation, access to resources were identified in the study. Overall, this study fills a gap in the literature on how these newer factors shape team productivity and project outcomes based on the success rate of the product the team developed.

*Keywords:* Constitutive Communication of Organizations (CCO), Product Design Teams, Product Management , Software Developer, User-Experience, Emerging Professional.

Curiosity, not experience, is a key factor for team success. When a Product Manager was asked what the key factor was, he replied "I think the biggest one is asking questions". In the world that is increasingly shaken by AI and uncertainty exists on job prospects, one key factor stands out for candidates: strong communication skills. Communication is defined as the exchange of information between entities under the potential influence of noise or interference (Shannon and Weaver Model of Communication, n.d.). This definition can be expanded into organizational communication which is defined as the exchange of information to guide, create and collaborate between organizational entities under the influence of noise or influence (Park University, 2025). In the context of an organization like a department that develops new products (i.e. New Product Design teams or NPD team) there are many entities involved in the development of the product. These entities are stakeholders. With these stakeholders, they co-create and collaborate in the organization. As this co-creation unfolds, there are noises that influence how that co-creation happens towards a product goal.

This whole process, especially the noises, has been studied from users to clients to product managers and developers (Sivasubramaniam, Liebowitz, & Lackman, 2012; Gallivan & Keil, 2003; Edmondson & Nembhard, 2009). However, there is less study done on the factors or noises attributed or encountered by a demographic, emerging professionals - who are young people between the age of 18-27 with less than five years of experience. These emerging professionals per 2025 projections are expected to comprise 27% of the global workforce by year-end, a figure echoed by the World Economic Forum's earlier estimates (Kelly, 2025).

Supported by the data, this study explores the potential of emerging professionals and the factors that they contribute to a New Product Design team setting. Guided by the theory of Communicative Constitution of Organizations, this study looked into what factors influenced the co-creation of a product's success in a New Product Design team. Building on this analysis, this study argues that teams composed of emerging professionals exhibit factors that differ from those of established professionals in New Product Design (NPD) contexts.

Literature Review

Every New Product Design team's ability to successfully produce a product depends heavily on how effective, efficient, and market ready the team is (Sivasubramaniam, Liebowitz, & Lackman, 2012, pp. 803, 807). However scholars note that market readiness is not straightforward. Paetzold and Stoiber (2025) argue that a product's success becomes questionable when users emphasize causality, meaning how they use the product from their perspective, while developers emphasize finality, meaning how functionality is prioritized (p. 2). These two orientations create a form of mutual ignorance in which causality and finality lead to misunderstanding between users and developers.

To address this mutual ignorance Paetzold and Stoiber (2025) examined how noise within user developer communication can be identified and reduced through prototyping (p. 2–3). However, existing studies overlooked a distinct factor within developer populations, namely the role of emerging professionals, defined as individuals aged 18 to 27 with fewer than five years of experience. Emerging professionals may have limited academic backgrounds, limited work experience, and limited social skills in workplace communication. As a result, mutual ignorance may manifest differently within teams composed of emerging professionals. This gap indicates a need to revisit user developer communication in prototyping from the perspective of emerging professionals.

Within these research studies, which reflect the needs of developers, the concept of psychological safety is a critical factor. This encourages developers to speak up, be comfortable, and address their concerns (Edmondson & Nembhard, 2009, p. 131). This sense of safety is shaped in part by leadership practices, particularly the degree to which leaders are communicative and accessible.  This safety is reflected in the findings, as higher productivity and quality are achieved when leaders communicate with the team either through providing safe space or having easy access to leadership guidance (Ehrlich & Cataldo, 2014, p. 741). However, much of this literature examines communication through strong, leadership-centered models. As a result, this gives limited insight into how centrality might function in non-conventional teams or how inclusive communication practices operate within less centralized team structures.

Even though there are studies on individual stakeholders, it is important to examine the entire communication chain, from users to developers. The factors that developers face are trickle-down effects from various communication interactions among stakeholders, particularly in User-Developer communication, which involves prototyping (Paetzold & Stoiber, 2025). Other factors and strategies exist beyond just one stakeholder communication within the form of the four-stage user-developer communication model, as evidenced by Gallivan and Kiel (2003). These factors include overconfidence bias (the biasness of the developer produced by their overconfidence on the solution they have developed) , communication breadth and depth (the variety of communication mediums and the deep understanding in each mediums), and fidelity (the accuracy of information that is communicated) (Gallivan & Kiel, 2003 p. 58-59). These factors are again focused on actual New Product Design teams, rather than New Product Design teams involving emerging professionals, which is why it is important to determine if these patterns and factors should be enhanced to fit New Product Design teams with emerging professionals.

Aside from the factors that exist for product success, as identified by product managers. There are factors attributed to established professionals who are also developers. These factors include a plethora of independent factors such as team tenure, functional diversity and more (Sivasubramaniam et al., 2012, p. 805). With factors such as team tenure, further research is needed to understand how participation patterns evolve because emerging professionals are balancing both school and work doing internships, co-ops and work-study programs. This means team tenure is very low. Because of this low team tenure, teams with emerging professionals are fluid in aspects of team bondings, teamworks, responsibilities and communications. Therefore, it is important to revisit these factors to verify if there are new factors.

Studies exist across many stakeholder relationships: users, clients, product managers, and developers. But this broad coverage of communication among stakeholders lacks one thing in common. Most of that literature focused on experienced professionals, but did not explicitly limit the research to emerging professionals. Accordingly, this research intends to fill the literature gap by asking: which factors uniquely associated with emerging professionals affect product success and performance in New Product Design (NPD) teams?

## Theoretical Perspective

The field of organizational communication has earlier viewed communication and organizations as two separate entities (Basque, Bencherki, & Kuhn, 2022, p. xxvii). By contrast, the Constitutive Communication of Organizations (CCO), contradicts the earlier assumption of communication and organizations as separate entities. As CCO theorists believe, communication calls organizations into being. (Basque, Bencherki, & Kuhn, 2022, p. xxxiv). This idea is understood as constitutive, which is the communication's capacity to create, maintain, and dissolve organizations (Basque, Bencherki, & Kuhn, 2022, p. xxvii). The second aspect is the idea that discourse and conversation are also the reasons for the creation of an organization (Basque, Bencherki, & Kuhn, 2022, p. xxviii). Out of these concepts come the three schools of CCO: the Montreal school, the Four Flows, and the Luhmannian School.

These three schools of CCO have many differences when it comes to how communication creates organizations. However, they all fall under the idea that communication constitutes organizing through space and time (Basque, Bencherki, & Kuhn, 2022, p. xxxi). The different concepts the three schools represent are as follows:

1. The Montreal school says organization constitutes through the formation of talks and texts (Cooren, 2004, p. 373);
2. The Four Flows school says organization constitutes through interactional function between stakeholders and their relationship with work, structure and institutional positioning (Basque, Bencherki, & Kuhn, 2022, pp. xxx–xxxii);
3. The Luhmannian school talks about closed communication where it self-produces for the organization. (Griffin, Ledbetter, & Sparks, 2023, p. 301)

Out of these three schools, the Four Flows school is applied in this research because it helps understand how interactional function affects the communication chain of the product team. In helping this research, the theory CCO's Four Flow school, founded by Robert McPhee, describes the way communication constitutes organizations as a series of four interactional flows (Griffin, Ledbetter, & Sparks, 2023, p 300 - 301). The first flow is membership negotiation, which McPhee describes as the exchange of ideas and setting boundaries between the organizations and its members (Griffin, Ledbetter, & Sparks, 2023, p 302). However, membership negotiation does not end when the member is accepted into the organization; they also need to maintain good relationships in the organization actively (i.e. socialization) (Griffin, Ledbetter, & Sparks, 2023, p 302). The second flow is the self-structuring of the organization. Self-structuring in the formal communication acts that create the organization. The third flow, activity coordination, focuses on how the organization gets things done, "as members communicate to adjust work processes or solve problems as they arise" (Griffin, Ledbetter, & Sparks, 2023, p 304). The fourth flow is institutional positioning, which is not within the organization, but it is how the organization portrays itself and represents itself among the myriad of networks, organizations, and people that exist (Griffin, Ledbetter, & Sparks, 2023, p 304 - 305).

     Closure is important as it exists throughout the four flows. Closure is when communication becomes stable and it starts to shape the organization. This closure appears through back-and-forth communication. With this closure, there is a component of time when the organization in the case of self-structuring needs approval or meet protests from past members or catering for future membership (Griffin, Ledbetter, & Sparks, 2023, p 303). Whereas an organization can also set goals and policies throughout its structure, which can be geographical, socio-political, what these communications create is consistency across these distributed spaces. It can be referred to as closure through space. Closure is important in this research because identifying those clusters is identical to finding factors that exist through space and time.

     With the theory, looking into the different flows on how organizations are constituted through communication, is integral in this research because all flows scrutinize how an

organization is formed from a discursive angle. As it is used to analyze the complete communication chain, the structure will be broken down and assigned based on the four flows. This analysis helps the research make sense of what a job environment with emerging professionals looks like, and how it can be improved to support product success by identifying relevant factors and strategies not seen within the established literature.

## Methods

The method used in this research is quantitative surveys and qualitative interviews with the research and methods gained approval from the IRB. The PI went to the product development environment where the emerging professionals study and work. And reached out to the software team with 10 student developers who worked with 4 clients in the area. The team heavily used ad hoc communication and relied on the students to deliver and update the products. The interviews and surveys are IRB approved. Both surveys and interviews are done through consent with further consent being taken for voice recordings. The research was made sure no one is below 18. The data that is collected is also protected and kept safe.

Some factors are identified that can be attributed to a product's success. This is achieved by involving all four different stakeholders through the communication chain of the software development:

1. Developers - Emerging professionals (ages 18 - 27) who performed software development work. Any identity is eligible, except for age restrictions.
2. Supervisors - Product managers and team leads who guided developers' tasks. Also unrestricted demographically except for age.
3. Users - Individuals who used the software; these may overlap with clients.

    4. Clients – Individuals who commissioned, requested, or interacted with the product and communicated with supervisors and occasionally developers.

A quantitative survey with demographic questions was administered across stakeholder groups, with survey focus varying by role. Users were surveyed on software usability and interaction experience (n = 38). Clients were surveyed on interactions with product managers and perceptions of communication effectiveness (n = 1). Developers and product managers were surveyed on workplace communication practices, including collaboration and organizational communication structures (developers: n = 4; product managers: n = 2).

Data were collected primarily through opportunistic sampling based on participants' roles and existing professional connections. To reduce recruitment bias, the PI recruited participants through multiple supervisors and client representatives (Tracy, 2025, p. 91). For users, random sampling was employed to prioritize breadth of product satisfaction feedback (Tracy, 2025, p. 90). In addition, the PI conducted observational fieldwork while embedded in the product design space, which allowed access to technical contexts but introduced potential limitations due to researcher proximity (Tracy, 2025, p. 148).

Semi structured interviews were conducted with developers (n = 3) and product managers (n = 3) to explore communication culture, information flow, and satisfaction. Interviews lasted approximately 30 minutes each, were conducted in comfortable settings or via Zoom as an accommodation, and were audio recorded with participant consent (Tracy, 2025, pp. 86, 170).

*Data Analysis*

Qualitative data were analyzed using a Phonetic Iterative Qualitative Data Analysis approach, in which data were coded into themes across multiple iterations (Tracy, 2025, p. 225). The coded data were then interpreted through the Constitutive Communication of Organization framework to identify factors and strategies relevant to the research question. Survey data were synthesized descriptively and used to support qualitative findings.

## Analysis

Using the Four Flows framework the analysis organizes the data into four main themes that explain how communication constitutes the product design team. These themes include professional identity formation through membership negotiation, documentation practices that enable emerging professionals to understand organizational codes and self structure, and the use of organizational tools such as Trello. In addition ad hoc communication shapes interaction within fluid teams, while activity coordination highlights communication modes and evolving strategies.

Together these factors form the communicative processes that constitute the organization and enable it to assume an institutional position when negotiating with clients. Through this analysis the systemic communication factors shaping New Product Design teams composed of emerging professionals are identified.

*1 Becoming a Team Member: Reality and Identity*

When developers first join the team, they must undergo a series of gatekeeping activities, including interviews, cover letters, team culture assessments, and orientations. During these membership negotiations, developers are evaluated based on specific qualities sought by the team. Project managers are typically the first to identify these needs and play a central role within New Product Design teams. Alongside pre-existing developers, they function as the primary gatekeepers of the team. However, once a developer becomes a formal member, membership negotiation does not end; instead, it continues through ongoing interactions between developers and product managers.

*1.1 Work Reality: Work Expectations and High Turnover*

When members join the group, they first undergo interviews and orientation. These processes are designed to integrate them into the work culture and establish expectations for future performance. These work expectations are evident when product managers describe their gatekeeping responsibilities in selecting student developers. As Product Manager pr PM 3 stated, "I usually pick middle-of-the-road students who I can see a place where they could grow; I can mentor them," highlighting the role of product managers as gatekeepers. While these gatekeepers look for specific qualities during membership negotiation, a critical constraint shapes their expectations: because members are emerging professionals, identifiable traits must supplement the work experience they have yet to acquire.

These role expectations serve as a preliminary indicator of readiness. As Participant 2 or P2 noted, "I had achieved and showed good academic results, which also can prove that I am eligible for the role," referencing both academic performance and the institutional requirement of completing "an internship during the freshman year summer, which allowed anyone to be in this position." Together, these forms of achievement signal capability in place of years of industry experience.

The prioritization of role expectations over work experience is driven by an underlying factor: **high turnover**. When emerging professionals participate in New Product Design teams, membership negotiation is continually shaped by the expectation that new members will regularly enter the team. Emerging-professional programs such as work-study colleges, internships, co-ops, and externships experience high turnover because they are designed to prioritize learning and work exposure rather than long-term task completion. Under these conditions, team tenure becomes a negligible factor, as emerging-professional teams operate with consistently low tenure by design. As PM 3 noted, "But every year, that whole team changes, right?" This point was further reinforced when they stated, "We can't build our understanding on prior knowledge because we get a new cohort who doesn't have any of that every single year" (Sivasubramaniam, Liebowitz, & Lackman, 2012, p. 805). Rather than evaluating team efficiency based on tenure, recognizing high turnover as a constant membership reality allows organizations to redirect resources toward mitigating its effects.

*1.2 Professional Identity Formations : Curiosity, Respect & Inclusion*

These organizations and their gatekeepers are selective about who they accept, based on the needs of the team. However, this process of membership acceptance introduces two forms of diversity: **identity diversity** and **functional team diversity**. Identity diversity is particularly important in workplaces where differences in identity and belief are explicitly valued. In this context, identity diversity can also become a structural consideration, as the demographic composition of the investigated software department reflects imbalances in representation. For example, the department historically maintained a **3:2** male-to-female ratio among primary student labor, but for one recent academic term, the ratio dropped to **1:0**.

As demographics shift, environments in which software developers are recruited must remain inclusive. This expectation is reflected in participant accounts. As P1, who identifies as Black or African American and as female, stated, "So I have not faced any in this workplace specifically, I have not faced any like discrimination because I am a minority or something like that." This sentiment is reinforced by another participant who also identifies as Black or African American and as female, noting, "When we come to it, I would be placed under a minority, but I feel like this is a very inclusive environment." These accounts indicate that inclusivity is essential for fostering effective communication within the organization. This finding reinforces the importance of psychological safety, which encourages developers to speak up, feel comfortable, and address concerns (Edmondson & Nembhard, 2009, p. 131). As such, membership negotiation should not jeopardize a member's core identity.

Apart from identity diversity, **functional team diversity** is sometimes associated with a "negative effect on team performance," as diversity has been linked to higher levels of dissatisfaction and turnover. However, research involving emerging professionals settings suggests a different dynamic, as their core values and expectations differ from those of established professionals (Edmondson & Nembhard, 2009). While Edmondson and Nembhard (2009) emphasize the importance of talent diversity, their discussion is primarily situated within contexts of established professional teams. As a result, the limited literature on emerging professionals, combined with the predominance of research focused on established teams, provides insufficient clarification of how functional diversity operates in emerging-professional settings. This gap is reflected in PM 3's observation:

> "Well, inexperience would be the biggest one (because) you all (emerging professionals) generate sloppy code because you are new to this, right?

> Moreover, it is not like a bash on the students. You are teaching it. I wrote bad code when I was your age, too."

Professional identity formation during membership negotiation does not depend primarily on an individual's technical talent; instead, it is shaped by two key themes. The first theme is **respect**, which emerges through identity negotiation between developers and the team. As P1 explained, "a professional way. So I do my best to like the relationship with everyone and like a relationship based on respect. I respect you, respect me, I communicate with you, you communicate with me." These reciprocal expectations of respect actively shape how professional identity is formed within the team.

The second theme is **curiosity**. Edmondson and Nembhard (2009) suggest that "team members must adopt an inquiry orientation in which they explain their position and inquire of others." This inquiry orientation includes asking questions of peers and team leads; however, within emerging-professional teams, it extends beyond formal questioning to encompass taking initiative in seeking answers and learning independently. As such, inquiry orientation alone is insufficient to capture this dynamic. Instead, **curiosity** emerges as a recurring and central factor throughout the data. This emphasis is reflected in PM 3's statement, "They have to be encouraged to ask questions," as well as P3's observation that effective developers demonstrate "good social skills, willingness to learn, communicating, (and) curiosity," along with what P1 stated:

> " that curiosity is actually a quality because curiosity made me learn, you know, because it's about asking those questions so that I am able to understand"

*2 How Work Is Organized: Documentation and Constraints*

When a member, regardless of PM or Dev, is accepted into the "organization," membership negotiation still exists, but self-structuring plays a significant role in how the team's rituals and customs shape the New Product Design team. These are closures, and over time, they influence the Agile methodologies and various development techniques a team will use.

*2.1 Documentation Practices*

When the organization communicates there is **strict documentation** and this has changed over time. Within the student team space there used to be a physical Kanban board. This tool was used to keep track of the status of multiple tasks. This system has remained intact but has evolved into a digital format. This structuring of information across time has enabled the organization to be more purposeful especially when the workforce comprises emerging professionals. Moreover documentation on how product design teams should work and interact as well as what to design and what not to choose can be standardized and recorded.

This process begins by associating new members with the tools the team uses as P3 stated that they "wish that for the internship especially for this program as well I wish there is like a certain course they want us to take like a YouTube playlist." This learning framework factor can be translated into **documentation** as a living document where multiple developers and product managers constantly contribute and correct. This form of documentation is helpful similar to Trello where emerging professionals wherever they are can digitally update their status across the board or reach out to a graduated senior to understand why a particular product was designed in a certain way. These communication mediums of **structural documentation** are not just guidance but a closure of time and space in the four flows. However an organization can become de structured when this **structural documentation** fails as PM 1 said:

> "Like, ideally, you have enough examples and guidance on what a UI should look like in the rest of your application. You can say, 'You know, we need a submit button and a notification for failure or something like that.'"

However this lack of **documentation depth** was reflected when PM 1 confirmed that it has not "maintained as often as it should be." This concern was reiterated in the average score of 7.5 for the question "How clear is your communication with product managers?" which included Trello documentation and GitHub. This is further reinforced by P3 who stated "The Trello cards sometimes are not descriptive in here" and by P1 who noted "the issue is not described properly but maybe there are not enough details."

These gaps in information can affect the **self structuring** of the organization and its ability to coordinate work as this **documentation** is clearly identified as necessary through the attributions of both product managers and participants. However this **documentation** is not a factor identified in existing literature. As such it becomes a noteworthy factor particularly for emerging professionals where adequate documentation is essential for role clarity coordination and overall team performance.

*2.2 Environmental Constraints*

While **documentation** exists to support the team it also helps maintain the workforce structure. However there are still constraints that have explicitly affected team productivity. As developers are emerging professionals the most basic constraint is a lack of experience. This lack of expertise is not treated as a setback but rather as a factor that must be accounted for. Because emerging professionals include student laborers, interns and new graduates there is an ongoing expectation that learning and growth are part of the role. This opportunity to gain experience coincides with another factor which is balancing studying and working as PM 3 stated:

> "Um, the second thing is you know you've got 10 hour contracts during the academic year. That is not a fourth of a 40 hour contract in reality. Somebody working 40 hours, you know eight hours a day focused on one problem, is not the same as a student having 10 hours spread across all of their classes and other things that they're doing as a student and being expected to do, you know, 10 hours of work. There's lots of **context switching** you have to get back into the problem you were doing after you've left it."

Moreover another factor identified is **location proximity**. Compared to what exists in established literature such as leadership centrality where information flows from the leader this factor expands on how **ad hoc communication** improves when the product manager is physically close. As P2 stated:

> "First with the supervisor I guess his office is always open during our work time. He's always there. So whenever a question pops up, any question I want to ask has always been there."

This sentiment is supported by P1 who stated that **direct communication** is their first choice when it is facilitated by proximity: "So usually like in terms of communication really for me it's like going towards the person. And if that doesn't work then texting the person."

These **environmental constraints** illustrate the conditions emerging professionals must balance in student labor positions and internship settings. Because of these factors it is essential to consider environmental constraints from the perspective of emerging professionals when evaluating team productivity and coordination.

*3 Doing the Work: Coordination and Communicating*

The organization, aside from its **structural documentation**, also relies on **ad hoc communication and mediated communication** to shape the organization beyond its physical space. In New Product Design teams these communication modalities include Teams, Slack, to passing by and talking. The ad hoc form of communication is uniquely used by emerging professionals.

This **ad hoc communication** supports emerging professionals when direct interaction happens unscheduled by location such as bumping in buildings, classes, or activities. Through communication across space the organization becomes unstructured beyond the four walls of the office. As PM 1 stated, "Okay, day to day with clients is usually not particularly scheduled, so it's an ad hoc communication."

*3.1 Ad Hoc Communication and Boundary Spanning*

Throughout the work environment a software team may choose multiple ways of structuring such as forms of **documentation**. However **material infrastructure** differs from spontaneous communication such as **ad hoc communication** and **boundary spanning** which involve unstructured interactions. This **ad hoc communication** occurs largely because developers face time constraints as they balance studying and working. As PM 3 stated "And so our customer can see something happen in a couple of days whereas our team it takes four times as long cause they're students and we're on 10 hours."

This dynamic can be observed in emerging professional teams that are not confined to cubicles or offices unlike established professional teams. When this occurs communication ranging from asking questions to requesting code reviews often becomes informal and in-person. Developers may drop by or drop in to speak directly with others as P1 stated:

> "I just go to the person and be like 'Hey are you free Can you please review this and that.' If they are not free then I try again with the other person. And if no one is there I usually put it on the board and be like this is my file and then they will [see it] when the place is empty."

Similar to the closure of space and time caused by time constraints is the developer ability of **context switching**. This factor directly affects **work coordination** as PM 3 explained:

> "That context switching takes time that eats into that 10 hours so having good processes and good ways of making sure we leave ourselves in a place where we can pick it up quickly the next day is important."

This **context switching** is only possible when supported by effective **documentation** that allows developers to reenter tasks quickly. These variables reflect the reality of **ad hoc** **communication** which is not addressed in literature that examines how "two particular team level communication structures hierarchical and small world patterns impacted product development team outcomes differently". Emerging professional teams rely primarily on **mediated communication** and **ad hoc communication** as their dominant team level communication structures. Hierarchical communication is weakened due to **location proximity** and accessibility factors (Ehrlich & Cataldo, 2014, p. 737).

*3.2 Miscommunication*

Within **ad hoc communication** and **context switching** the unstructured nature of these factors can lead to **miscommunication**. When developers first receive instructions they may struggle to determine what needs to be implemented because issue descriptions are short or non descriptive. This often results in back and forth communication or ping pong communication between the product manager and the developer where clarification is required.

After clarifying the issue the developer must often reach out again to describe their implementation and identify potential edge cases as P1 stated "and if along the way there's still a misunderstanding then I will check again because I usually like to check as I'm going". These **miscommunications** occur due to noise in **ad hoc communication** that lacks sufficient **documentation**. Valuable information may be buried among excessive messages or not provided in enough detail which contributes to this noise.

Despite these challenges survey results indicate that **location proximity** and access to information sources help mitigate miscommunication. When developers were asked "How often are misunderstandings with product managers resolved effectively" they reported a satisfaction rate of 8.5 out of 10. Similarly when asked "How helpful is the feedback you receive from product managers" developers reported a satisfaction rate of 8.25 out of 10.

*4 Bridging Worlds: Managing Client–Team Expectations*

Through **activity coordination**, **self structuring**, and **membership negotiation**, communication within the team constitutes the organization. However this focus on internal communication leaves one aspect unaddressed. External communication also plays an essential role particularly when developer teams interact with clients. As client needs are translated into problem statements for developers to solve.

*4.1 Product Manager as mediator on Expectations and Development Realities*

As the PM facilitates talks between the developers and product managers, they become mediators. As the developers are emerging professionals, they may struggle to empathize with their users and clients. As the constraints of developers are experienced, the need for a captain at the helm of the ship to deliver goods is necessary, as PM 3 analogizes :

> "I would say now, and then you can also tell what you've done in that. Okay, so right now, my role is very much I am more like an admiral of a ship than it is, like captain or first mate or anything like that. So I really am in charge of seeing the big picture, the fleet of our software, how is it all moving forward?"

When this is the role, the product managers are tasked with balancing the client's expectations first. This means that during the meetings, the client will present the functionalities or new implementations they want to implement. The product manager needs to recognize the development realities of the developer team, which means understanding the time, experience, and numerical constraints that the team has, as stated by PM 3:

> "In those meetings, we can say, 'That's a really cool idea. I don't think our team can implement that because of,' you know, in the list of limitations, so it's an experience thing. So I can say pretty easily, 'Oh, we're not building a social platform for you.' "

Only then can product managers promise clients the timeline they can deliver, as delivery is part of building trust, as both PM 1 and PM 3 state: "so you build trust by listening to them and delivering what you've said you will." When the product manager can recognize the team's constraints, they can project what the organization's limitations are. This does not mean the organization is dysfunctional, but when it fails to meet goals and set expectations that cannot be fulfilled, it breaks the trust between the team and the client.

*4.2 Visual and Demonstrative Alignment in Feedback Loops and Interaction Channel*

The other time when product managers typically meet with clients is when a particular feature or implementation requires an update or needs to be reviewed for approval. During this time it is essential that the developer's work reflects the problem statement received by the client. When showcasing the team's work product teams often rely on **visual communication** to bridge the gap between technical terms, team processes and the client who may not be familiar with them. As PM 2 stated "We pull up the screen with all the changes that we have done and usually me and my manager update the clients with the changes that we have done."

This approach helps bridge potential miscommunication as developers must package the implemented feature into something visually understandable. As PM 1 stated:

> " It's a visual thing where instead of both of you trying to convert your thoughts into the language and then understand the language, you can actually show this is what we're expecting or this is what we have heard from you."

In this way, this removed the noises that the dev team wanted to present in a way that made them understandable. This strategy, in turn, reinforces Paetzold and Stoiber (2025) on how prototyping reduces the mutual ignorance that often occurs between users and developers.

## Conclusion

With the studies and multiple interviews, there are certain factors and strategies deployed by the product design teams that lead to its product success rate being rated by its users as 7.81 out of 10, while the user's satisfaction rate is 7.51 out of 10, which implies that the factors identified have contributed to the success of the product as it comes to emerging professionals in the New Product Design team. This research addresses a gap in existing literature by identifying factors uniquely associated with emerging professionals that affect product success and performance in New Product Design teams. Through the CCO framework the analysis reveals that membership negotiation in learning oriented teams prioritizes curiosity, respect, and expectations rather than prior experience. While the scope of the study is limited to a single team the findings highlight factors not emphasized in established literature, including documentations, environmental constraints, context switching and ad hoc communications.

Talks that Builds
24At the same time the analysis affirms existing factors such as psychological safety while showing that others do not fully apply due to high turnover and functional diversity in emerging professional teams. Together these findings help clarify what a job environment with emerging professionals looks like and how organizational communication can be improved to better support product success.

Data Table of User Satisfaction for Product Success

| Questions | Mean (0-10) | Standard Deviation | Responses |
| --- | --- | --- | --- |
| How satisfied are you with the product overall? | 7.51 | 2.17 | 35 |
| How easy is it to use the product? | 7.64 | 2.19 | 36 |
| How well does the product meet your needs? | 7.86 | 2.19 | 36 |
| How reliable is the product (few errors, consistent performance)? | 7.89 | 2.00 | 36 |
| How satisfied are you with the product's design/interface? | 7.58 | 2.18 | 36 |
| How effective is the product in helping you complete your tasks? | 7.47 | 2.51 | 36 |
| How well does the product save you time or effort? | 7.44 | 2.61 | 36 |
| How satisfied are you with the product's accessibility (availability, ease of access)? | 7.33 | 2.45 | 36 |
| How well does the product compare to alternatives (if you've used others)? | 7.54 | 2.12 | 35 |
| How well does the product handle updates or improvements? | 7.49 | 2.18 | 35 |
| How responsive are support or help resources when you face issues? | 7.64 | 2.18 | 36 |
| How much trust do you have in this product? | 7.47 | 2.51 | 36 |
| How likely are you to continue using this product in the future? | 7.50 | 2.84 | 36 |
| How likely are you to recommend this product to others? | 7.58 | 2.55 | 36 |
| How successful do you consider this product overall? | 7.81 | 2.31 | 36 |

Reference:

Basque, J., Bencherki, N., & Kuhn, T. (Eds.). (2022). The Routledge handbook of the communicative constitution of organization. Routledge. https://doi.org/10.4324/9781003224914

Bonito, Keyton, & Ervin (2017) Bonito, J. A., Keyton, J., & Ervin, J. N. (2017). Role-related participation in product design teams: Individual- and group-level trends. Communication Research, 44(2), 263–286. doi:10.1177/0093650215618759

Cooren, F. (2004). Textual agency: How texts do things in organizational settings. Organization, 11(3), 373–393. https://doi.org/10.1177/1350508404041998

Edmondson, A. C., & Nembhard, I. M. (2009). Product Development and Learning in Project Teams: The Challenges Are the Benefits. Journal of Product Innovation Management, 26(2), 123–138.

Ehrlich, K., & Cataldo, M. (2014). The Communication Patterns of Technical Leaders: Impact on Product Development Team Performance. In Proceedings of the ACM Conference on Computer Supported Cooperative Work and Social Computing (CSCW '14), (pp. 15–19).

Gallivan & Keil (2003) Gallivan, M. J., & Keil, M. (2003). The user–developer communication process: A critical case study. Information Systems Journal, 13, 37–68.

Griffin, E., Ledbetter, A., & Sparks, G. (2023). *A first look at communication theory* (11th ed.). McGraw-Hill Education

Kelly, J. (2025, April 1). *Gen-Z's are redefining the way they want to work*. Forbes. https://www.forbes.com/sites/jackkelly/2025/04/01/gen-zs-takeover-and-redefining-the-workplace/

Paetzold, K., & Stoiber, L. (2025). Development of a communication model for the efficient exchange of information between user and designer. Proceedings of the Design Society, 5: ICED25. https://doi.org/10.1017/pds.2025.10150

Park University. (2025, June 27). What is organizational communication? An introduction. https://www.park.edu/blog/what-is-organizational-communication-an-introduction/

Shannon and Weaver Model of Communication. (n.d.). Communication Theory. https://www.communicationtheory.org/shannon-and-weaver-model-of-communication/

Sivasubramaniam, Liebowitz, & Lackman (2012) Sivasubramaniam, N., Liebowitz, S. J., & Lackman, C. L. (2012). Determinants of new product development team performance: A meta-analytic review. Journal of Product Innovation Management, 29(5), 803–820. doi:10.1111/j.1540-5885.2012.00940.x

Tracy, S. J. (2025). *Qualitative research methods: Collecting evidence, crafting analysis, communicating impact* (3rd ed.). John Wiley & Sons, Inc.

Wing, Petkov, & Andrew (2020) Wing, J. W., Petkov, D., & Andrew, T. N. (2020). A systemic framework for facilitating better client-developer collaboration in complex projects. International Journal of Information Technologies and Systems Approach, 13(1). doi:10.4018/IJITSA.2020010103